# Size diversity of old Large Magellanic Cloud clusters as determined by internal dynamical evolution

F.R. Ferraro[1,2*], B. Lanzoni[1,2] , E. Dalessandro[2], M. Cadelano[1,2], S. Raso[1,2], A. Mucciarelli[1,2], G. Beccari[3], C. Pallanca[1,2]

[1]*Dipartimento di Fisica e Astronomia, Universita` di Bologna, Via Gobetti 93/2, I-40129 Bologna, Italy*

[2]*INAF -- Astrophysics and Space Science Observatory Bologna, Via Gobetti 93/3, I-40129 Bologna, Italy*

[3]*European Southern Observatory, Karl-Schwarzschild-Strasse 2, 85748 Garching bei Mu□nchen, Germany*

**The distribution of size as a function of age observed for star clusters in the Large Magellanic Cloud (LMC) is very puzzling: young clusters are all compact, while the oldest systems show both small and large sizes. It is commonly interpreted as due to a population of binary black holes driving a progressive expansion of cluster cores. Here we propose, instead, that it is the natural consequence of the fact that only relatively low-mass clusters have formed in the last ~3 Gyr in the LMC and only the most compact systems survived and are observable. The spread in size displayed by the oldest (and most massive) clusters, instead, can be explained in terms of initial conditions and internal dynamical evolution. To quantitatively explore the role of the latter, we selected a sample of five coeval and old LMC clusters with different sizes, and we estimated their dynamical age from the level of central segregation of blue straggler stars (the so-called dynamical clock). Similarly to what found in the Milky Way, we indeed measure different levels of dynamical evolution among the selected coeval clusters, with large-core systems being dynamically younger than those with small size. This behaviour is fully consistent with what expected from internal dynamical evolution processes over timescales mainly set by the structure of each system at formation.**

The Large Magellanic Cloud (LMC) hosts star clusters covering a wide range of ages (from a few million, to several billion years), at odds with the Milky Way where mostly old (t>10 Gyr) globular cluster (GCs) are found. The LMC thus offers a unique opportunity to explore the evolutionary processes of stellar clusters over cosmic time. One of the most intriguing features emerging from these studies[1-4] is the behaviour of the core radius ($r_c$, which characterizes the size of the innermost cluster region) as a function of age (as measured from the cluster stellar population): the youngest clusters are all compact (with $r_c$ < 2.5 pc), while the oldest ones span the full range of observed $r_c$ values, from a fraction of a parsec to almost 10 pc (Figure 1, panel a), similar to what is measured for the Milky Way (old) GCs. After ruling out any possible bias due to selection effects, the observed trend has been interpreted in terms of an evolutionary sequence[4]. In this scenario, all clusters formed with compact cores ($r_c$ ~2-3 pc), then most of them maintained small cores, while several others experienced core expansion and moved to the upper-right corner of the diagram. Such an expansion, however, needs to be powered by some "ad hoc" mechanism. Among the different possibilities discussed in the literature[5,6], one often quoted scenario[7] is that the core expansion is due to the heating action of a population of stellar-mass binary black holes (BHs) retained after the supernova explosions. Dynamical interactions among single and binary BHs led to multiple BH scatterings and ejections, thus driving the expansion of the central cluster regions.

**An alternative reading of the cluster size-age distribution**

Although intriguing, the proposed scenario implicitly requires an evolutionary link between the younger and the older GCs in the LMC, with the former being representative of the progenitors of the oldest ones. However, the two groups show different masses and positions within the LMC: all the young clusters are light stellar systems (with $M<10^5$ $M_\odot$), while old clusters are all more massive than $10^5$ $M_\odot$ (panel b in Figure 1); moreover, the young objects are observed in the innermost regions of the host galaxy (within ~5 kpc from the centre), while the old ones are orbiting at any distance (panel c in Figure 1). These pieces of evidence strongly indicate that the progenitors of the old LMC clusters must have been more massive (up to a factor of 100) than the currently young systems, hence there does not appear to be a direct evolutionary connection between the two groups. In turn, this seriously challenges the reading of the LMC $r_c$-age distribution in terms of an "evolutionary sequence".

On the other hand, the observed distributions (Figure 1) show how the cluster parameters changed over the time in the LMC:

- during the early formation epoch of the LMC (~ 13 Gyr ago), many star clusters more massive than $10^5$ $M_\odot$ formed over a quite short time scale (the old clusters, in fact, all formed within a period of ~ 1 Gyr – see Methods section 'The age of the five LMC clusters') at any distance within the galaxy;
- after a long period of quiescence ($\Delta t$ ~10 Gyrs, the so-called "age-gap")[8-10], about 3 Gyr ago cluster formation was reactivated (likely because of a strong tidal interaction with the Small Magellanic Cloud)[11] and only less massive structures have been generated since then (i.e., over a much more extended period, of several Gyrs) essentially in the innermost region of the galaxy (Rg< 4-5 kpc) around the LMC bar[11].

Within this scenario, the lack of young clusters with large $r_c$ would be the natural consequence of the observed mass-age and distance-age distributions: since all recent clusters are light systems formed in the innermost region of the LMC, only the most compact ones can survive the tidal effects of the host galaxy, while any loose and light system that might have formed had been already disrupted. This directly explains why the upper-left portion of the $r_c$-age diagram is empty. According to the observed mass distribution of the old clusters, none of the young light systems currently observable in the LMC will probably survive over the next 10 Gyr.

Following these considerations, it remains to be understood why old GCs span a wide range of rc values. Here we propose that this is primarily due to a combination of different properties at the moment of cluster formation and different stages of internal dynamical evolution (different dynamical ages) currently reached by each system, with the larger-core GCs being dynamically less evolved (younger) than those with small $r_c$. Indeed, it is well known that GCs are dynamically active stellar systems, where gravitational interactions among stars can significantly alter the overall energy budget and lead to a progressive internal dynamical evolution[12] through processes like mass segregation, evaporation of light stars, core collapse, etc. Thus, star clusters formed at the same cosmic time (i.e., with the same chronological age) may have reached quite different stages of dynamical evolution, corresponding to different modifications of their internal structure with respect to the initial conditions. An innovative method to empirically measure the level of dynamical evolution suffered by a stellar system has been recently proposed[13-15] based on blue straggler stars (BSSs). These peculiar objects are thought to be generated by some mass-enhancement processes, like mass-transfer in binary systems[16] or stellar mergers due to direct collisions[17]. Independently of their formation mechanism, BSSs constitute a population of heavy objects (with average masses of 1.2-1.3 $M_\odot$ in old stellar systems)[18,19] orbiting a sea of light stars (with an average mass $<m> \sim 0.3$-$0.4$ $M_\odot$). For this reason they suffer the effect of dynamical friction, that makes them sink toward the cluster centre. The progressive (dynamical friction induced) central segregation of BSSs is very efficiently described[20] by the parameter A+, defined as the area enclosed between the cumulative radial distribution of BSSs and that of a reference (lighter) population (typically with a mass of ~0.8 $M_\odot$): N-body[20] and Monte Carlo[21] simulations demonstrated that, as expected for a sensitive indicator of the BSS sedimentation process, the value of A+ systematically increases with time. From an empirical point of view, this parameter has been measured in a large sample of old Galactic GCs[14,15], finding a strong correlation with the core relaxation time ($t_{rc}$) and thus confirming that the level of BSS central segregation is a powerful indicator of the dynamical age of the parent cluster: the method is therefore dubbed[13] the “dynamical clock”. Following this line of reasoning, we propose that the $r_c$ spread observed for old GCs in the LMC could be explained in terms of different levels of dynamical evolution reached by systems of fixed chronological age.

**The dynamical ages of five old star clusters in the LMC**

To provide arguments to support this scenario, we determined (through the $A_+$ parameter) the dynamical age of a sample of old LMC clusters, for which Hubble Space Telescope observations adequate enough to properly study the BSS population and reliably evaluate the LMC field star contamination are available (see Methods section ’The Data set’). The 5 selected clusters (namely NGC 1466, NGC 1841, NGC 2210, NGC 2257 and Hodge 11) are old, massive ($\log M/M_\odot \sim 5.2\pm0.2$)[22] and metal poor ([Fe/H]$\sim -1.9\pm0.2$)[22]: they are marked with red squares in Figure 1. The photometric catalogues were first used to re-determine the gravitational centre of each system. In fact, a correct location of the cluster centre is a key step, especially in such distant stellar systems, since even small errors can significantly affect the derived radial behaviour of the observed stellar populations. With respect to previous works we found differences up to several arcseconds (see Table 1). Note that one arcsecond corresponds to 0.24 pc at the distance of the LMC (we assumed d=50 kpc)[23]. We then determined new star density profiles and structural parameters (namely the core, half-mass and tidal radii, the concentration parameter, etc.) from resolved star counts and by properly taking into account the LMC field contamination (see Methods section ‘Field Decontamination’ and Table S1). According to similar works[15,24] performed on Galactic GCs, the BSS population has been selected in the "normalized Colour Magnitude Diagram (n-CMD)", where the magnitudes of all the measured stars are shifted to assign coordinates (0,0) to the colour and magnitude of the Main Sequence Turn-off (MS-TO) point. The co-added n-CMD of the 5 target clusters is plotted in Figure 2 (grey dots): as apparent, the main stellar evolutionary sequences of the five GCs are remarkably well superposed one on another, suggesting that these systems are all coeval. Moreover the perfect match with the CMD of M30, one of the oldest Milky Way GCs with comparable metallicity[25], suggests a common age of ~13 Gyr, (see Methods section ‘The age of the five LMC clusters’). Hence, the BSS population has been identified by adopting the same selection box in all the target clusters. The same holds for the selection of the reference population, i.e., a sample of normal cluster stars tracing the overall star density profile of the system. In particular, to be consistent with previous works performed in Galactic GCs[15,24]:

(1) We only considered BSSs with normalized V magnitude $V^* < -0.6$. This selection includes only the most massive portion of the BSS population, thus maximizing the sensitivity of the A+ parameter to the dynamical friction effect. Moreover, it excludes the faintest portion where increasing photometric errors and blends can make the BSS selection more problematic.

(2) As reference population we adopted the lower portion of the Red Giant Branch and the Sub Giant Branch, in the same range of magnitudes of the selected BSSs. This indeed provides the ideal reference population, as it includes several hundred stars (thus making statistical fluctuations negligible), and it assures the same level of completeness of the BSS sample.

(3) We measured the A+ parameter within one half-mass radius ($r_h$). This assumption allows a direct comparison among the five different systems and with the large sample of Galactic GCs studied in the literature[14,15].

The n-CMDs for all the stars measured within one $r_h$ in the five programme clusters are shown in Figure 3, where the selection boxes of the BSS and REF populations are also drawn. The cumulative radial distributions of the two populations are plotted in Figure 4, where we also marked the measured values of A+ and related errors (see Methods section 'Errors in the measure of A+'). We find that NGC 1841 and Hodge 11 show a low level of BSS segregation ($A_+$ = 0.02-0.04), suggesting that they both are dynamically young, while NGC 2257, NGC 1466 and NGC 2210 have increasing values of $A_+$ (up to 0.24), corresponding to a moderate/large dynamical evolution. Hence, in spite of their comparable chronological ages, these systems show different levels of BSS segregation and, thus, different levels of dynamical evolution. This is further confirmed by the left panel of Figure 5, showing the dynamical age of the investigated LMC GCs (large red squares) as a function of A+. The dynamical age is expressed in terms of the ratio $N_{relax}$ between the chronological age of the systems (t=13 Gyr) and their current central relaxation times (trc, see Methods section 'Central relaxation times'). This index quantifies the number of trc occurred since the epoch of cluster formation: a large value of $N_{relax}$ means a dynamically evolved stellar system, while a small value means a dynamically unevolved cluster. As apparent, the BSS segregation level measured in NGC 2210 indicates that ~100 central relaxation times have occurred since the formation epoch of this cluster, while this index drops to 20-30 for NGC 1466 and NGC 2257, and to just a few units for Hodge 11 and NGC1841. The figure also shows an impressive match between the results obtained here for the five LMC clusters and those previously found for a sample of 48 old and coeval Galactic GCs (grey circles)[15], demonstrating that the "dynamical clock" can be efficiently used in any stellar environment. The right panel of Figure 5 shows the effect of dynamical evolution on the core size for the entire sample of 48 Galactic GCs (grey dots) and the five LMC systems studied here (large red squares). As can be seen from the nice correlation, clusters with large core radius are dynamically younger (with lower values of the A+ parameter) than compact systems. The former have possibly maintained unchanged or only slightly modified their initial structure (in terms of core size, concentration, central density), while all dynamically old clusters currently appear as quite compact objects, although they possibly formed with a larger core. Hence, the internal dynamical evolution tends to generate compact clusters, systematically moving large-core systems toward small-size compact objects over a timescale that mainly depends on the cluster structure. Panels (a) and (c) in Figure 1 suggest that also the local environment might have had some impact on the cluster dynamical evolution: in fact, the old GCs with smallest core radii are located at the smallest galactocentric distances, indicating that their internal evolution has been accelerated by an increased evaporation/tidal stripping of low mass stars in the innermost region of the LMC. Of course also the fraction of dark remnants (as BHs and neutron stars) retained within each cluster and their ejection timescale has an impact on the dynamical evolution of the system (in the sense of slowing it down for increasing retention fraction)[20]. However both these quantities are unknown at the moment, and only a few observational evidence of BH candidates in GCs has been found so far[26,27]. Hence, here we consider the action of dark remnants as a second-order effect on the dynamical cluster aging.

**Conclusion and future perspectives**

On the basis of these results, we conclude that the observed spread of $r_c$ at a given chronological age can be interpreted as the "natural" consequence of GC internal dynamical evolution that brings systems with relaxation time significantly shorter than their age to populate the small core radius region of the diagram. It is also somehow "natural" that chronologically old GCs display the largest spread of core sizes, since in this case a variety of initial configurations (with intermediate/short relaxation times) could have evolved toward small $r_c$ configuration. Of course the proposed scenario leaves completely unaffected the portion of the diagram corresponding to small chronological ages ($t=10^7$-$10^8$ yr), because all young clusters have relaxation times comparable to (or larger than) their age and their internal dynamical evolution processes had not enough time to move them toward the small $r_c$ portion of the diagram. It will be interesting to extend this study to the intermediate-age clusters (logt>8-9), which could also show evidence of different levels of dynamical evolution. Indeed a first attempt[28] to measure the dynamical age of 7 LMC clusters in this age range suggests quite modest levels of dynamical evolution. However a detailed analysis of the oldest clusters in this age range (with ages larger than 2-3 Gyrs, as NGC 2121, NGC 2155 and SL 663) is still lacking and it can certainly provide further hints on this topic.

The evidence presented in this paper provides a new interpretative scenario for the age-size distribution in the LMC clusters that does not require the action of BHs, but it is essentially driven by the cluster internal dynamical evolution. This scenario removes the necessity of an evolutionary path in which compact young clusters evolve into old globulars with a wide range of radii. Moreover, it provides further support to the other structural (see Figure 1) and chemical[29-31] pieces of evidence that already challenged such an evolutionary connection. Hence, this result redirects our attention to the cluster formation history in the LMC, its dramatic changes over the cosmic time and the environmental conditions at which this process is occurring.

# Methods

**The Data-set:** For this study we used a set of high-resolution images acquired with the Wide Field Channel of the Advanced Camera for Survey (ACS/WFC) on board the Hubble Space Telescope, secured under proposal GO14164 (PI: Sarajedini). We used the images acquired through the filters F606W (V) and F814W (I) to sample the cluster population, and those (typically located 5' from the cluster centre) obtained through the filters F606W and F435W (B) to sample the Large Magellanic Cloud (LMC) field population. For both data-sets, an appropriate dither pattern of a few arcseconds has been adopted in each pointing in order to fill the inter-chip gaps and avoid spurious effects due to bad pixels. The photometric analysis was performed via the point-spread function (PSF) fitting method, by using DAOPHOT IV[32], following the "standard" approach used in previous works[33,34]. Briefly, PSF models were derived for each image and chip by using some dozens of stars, and then applied to all the sources with flux peaks at least 3σ above the local background. A master list including stars detected in at least four images was then created. At the position of each star in the master-list, a fit was forced with DAOPHOT/ALLFRAME[35] in each frame. For each star thus recovered, multiple magnitude estimates obtained in each chip with the same filter were homogenised by using DAOMATCH and DAOMASTER, and their weighted mean and standard deviation were finally adopted as star magnitude and photometric error. Instrumental magnitudes were calibrated onto the VEGAMAG photometric system[36] by using the recipes and zero-points reported in the HST web-sites. Instrumental coordinates were first corrected for geometric distortions by using the most updated Distortion Correction Tables IDCTAB provided on the dedicated page of the Space Telescope Science Institute for the ACS/WFC images. Then, they were reported to the absolute coordinate system (α, δ) as defined by the World Coordinate System by using the stars in common with the publicly available Gaia DR2 catalog[37]. The resulting 1σ astrometric accuracy is typically ≤ 0.1 mas.

**The age of the five LMC clusters** – The CMDs obtained in the present work allowed us to tightly constrain the age of the five target clusters, which are considered as old stellar systems (with ages larger than 10 Gyr) in all the compilations present in the literature.

The co-added n-CMDs of the 5 targets are shown in Figure 2 (grey dots), where only stars within the half-mass radius of each system have been plotted to better highlight the cluster populations. As can be appreciated, the match among the main evolutionary sequences is impressive: the co-added n-CMD appears as a single population, thus demonstrating that the 5 clusters are indeed coeval within less than 1 Gyr. To quantify their age, we superimposed the n-CMD of M30[38], a Galactic GC with comparable metallicity ([Fe/H]=−1.9)[25] and very well constrained age (13 Gyr)[39,40]. Another impressive match of the main evolutionary sequences is found, demonstrating that M30 is coeval to the 5 LMC clusters. Thus an age of 13 Gyr±1 Gyr (with a conservative estimate of the error) can be assumed for the 5 LMC clusters.

**Cluster structural parameters** – Many papers in the literature[41-43] underline the advantages of using star counts, instead of surface brightness, profiles to derive the cluster structural parameters. In fact, surface brightness profiles are known to suffer from possible biases due to the presence of a few very bright stars, which instead do not affect the number density profiles. In spite of this, most of the morphological parameter estimates (including those for the 5 LMC clusters considered here) are still based on surface brightness profiles. We thus performed new determinations based on star count profiles. The full analysis, including artificial star experiments for the photometric completeness estimate, is described and discussed elsewhere. Here we just summarize its main steps and the structural parameters relevant for the present discussion. According to the procedure adopted in previous works[44,45], we first determined the centre of gravity (Cgrav) of each system by averaging the right ascension (α) and declination (δ) of all stars brighter than a given threshold magnitude (to avoid incompleteness biases) and lying within a circle of radius r. For the five clusters discussed here, the threshold magnitude is around the main sequence turnoff level, while the typical radius r varies from 6"-65" depending on the cluster morphology. The derived values of Cgrav differ by ~2"-3" from previous determinations, but for NGC 2257 where the difference amounts to almost 6" (see Table 1). To build the number density profile, we thus divided the photometric sample in (typically 15-20) concentric annuli centred on Cgrav, each one split into an adequate number of sub-sectors. The number of stars lying within each sub-sector (and with magnitude above a threshold adopted to avoid incompleteness biases) was then counted, and the star surface density was obtained by dividing these values by the corresponding sub-sector area. The stellar density in each annulus was then obtained as the average of the sub-sector densities, and the standard deviation was adopted as the uncertainty. The LMC background level has been estimated from the parallel observations, typically located at 5' from each cluster (see the section 'The data set'). These have the F606W filter in common with the cluster observations, thus allowing a consistent estimate of the level of LMC field contamination at any fixed magnitude limit. Once estimated, the background level was subtracted to the stellar density measured in each annulus, thus to obtain the density profile of the cluster. Finally, this has been compared with the family of King models[46] characterized by different values of the dimensionless parameter $W_0$, which is proportional to the gravitational potential at the center of the system. The best-fit solution has been determined through a procedure that minimizes the sum of the unweighted squares of the residuals and evaluates the corresponding reduced $\chi^2$. The uncertainties on the derived structural parameters have been estimated in agreement with other studies in the litterature[22,42]: they correspond to the maximum variations of the parameter within the subset of models that provide a $\chi^2_{\rm min} \leq \chi^2_{\rm best} + 1$, where $\chi^2_{\rm best}$ is the best-fit $\chi^2$, while $\chi^2_{\rm min}$ is the minimum $\chi^2$ obtained for every value of $W_0$ explored. The core and half-mass radii of the 5 clusters, which are relevant for the present discussion, are listed in Table 1, while the full discussion of the adopted procedure and the results obtained will be given in a forthcoming paper.

**Central relaxation time** – Central relaxation times have been computed by adopting the newly determined structural parameters and following the well-know relation[47]

$$t_{rc} = 8.338 \times 10^6 \times \frac{\sqrt{\rho_c}}{\ln(0.4 \times M_{cl} / m_*)} \times \frac{r_c^3}{m_*} yr$$

where $\rho_c$ is the central mass density in $M_\odot/pc^3$, $M_{cl}$ is the cluster mass in $M_\odot$; $m_*$ is the average star mass (here we adopted 0.3 $M_\odot$) and $r_c$ is the core radius in pc. The values of the central relaxation times for the five clusters are listed in Table 1.

**Field decontamination -** It is well known that the CMD of the LMC clusters can be significantly contaminated by field star interlopers observed along the line of sight. Unfortunately, given the LMC distance, a detailed separation between field and cluster stars based on proper motions is possible only for a few cases. Moreover, accurate Gaia DR2 proper motions are available only for the brightest stars. As a consequence, to asses the impact of field contamination in the five cases discussed here we used a statistical approach based on the comparison between the CMD stellar distribution observed in the innermost regions of each cluster and that of a region representative of the surrounding LMC field.

To this end, we accurately analysed all the available observations in the vicinity of the program clusters. For three of them (namely NGC 1466, NGC 1841 and NGC 2257) the field contamination turns out to be negligible, with only a few stars measured over the entire field of view (11 square arcmin) of the ACS/WFC parallel observations sampling the nearby LMC field.

In the case of NGC 2210 and Hodge 11, the LMC field appears to be more pronounced and we thus performed a statistical decontamination procedure. This required us to transform the boxes used for the population selection (see Figure 3) into the (V, B-V) plane, because the fields surrounding these two clusters have been observed in the F606W (V) and F435W (B) filters. The former is exactly the same filter used in the cluster pointings. Hence the magnitude range along the vertical axis is precisely anchored. To determine the (B-V) limits of the adopted selection boxes we made use of theoretical models[48] of the appropriate metallicity, coloured in the ACS/WFC filter system. The BSS selection boxes transformed into the (V, B-V) CMD are shown in Supplementary Figure 1. We then counted the number of stars in the parallel observations falling within the selection boxes and we determined the LMC field density dividing this number by the area of the ACS/WFC field of view. The number of field stars contaminating the BSS population is thus obtained as the product between the number of selected BSSs and the field density, and the same holds for the reference population. It is important to remind that in our analysis we are considering only the stars observed within one half-mass radius of each cluster. This corresponds to an area of only ~1 sq. arcmin (i.e., 1/10 of the ACS field of view) in the case of Hodge 11 (for which rh=36.3"), and just ~0.2 sq. arcmin (i.e., 1/50 of the ACS area) in the case of NGC 2210 (having rh=15.9"). The results are that the LMC field contamination to the reference population is completely negligible in both clusters, while a few selected BSSs are likely field interlopers. Their exact number is listed in Supplementary Table 1 for three radial bins ($r<r_c$, $r_c<r<r_h/2$, $r_h/2<r<r_h$) adopted to preserve the radial information. To determine reliable (i.e., field-decontaminated) values of A+ we thus randomly removed these numbers of stars from the BSS population sampled in each bin, and we repeated this random decontamination procedure 5000 times, each time registering the resulting value of A+. Supplementary Figure 2 shows the histogram of the obtained values. As can be seen, a peaked distribution with a small dispersion (smaller than 0.01) is obtained in both cases, thus testifying that the value of A+ is solidly estimated also in these contaminated clusters.

**Errors in the measure of A+** – As discussed in previous papers[14,15] the main source of errors in the determination of the parameter A+ is due to the relative small statistics of the BSS sample detected in each cluster. Thus, the uncertainties in A+ have been estimated by applying a jackknife bootstrapping technique[49].

Following this approach, given a sample of *N* BSSs, A+ is recomputed *N* times from samples of *N-1* BSSs obtained by excluding, each time, one different star. Thus the procedure yields *N* estimates of A+ and the final uncertainty on A+ is obtained as $\sigma_{A+} = \sqrt{N-1}\,\sigma_{\rm distr}$ , where $\sigma_{\rm distr}$ is the standard deviation of the A+ distribution derived from the *N* realizations. The uncertainties are listed in Table 1 and reported in each panel of Figure 4.

**Data Availability:** The data that support the plots within this paper and other findings of this study are available from the corresponding author upon reasonable request.

**Correspondence to:** F.R.Ferraro[1] Correspondence and requests for materials should be addressed to F.R.F. **francesco.ferraro3@unibo.it**

**Acknowledgements:** This research is part of the project COSMIC-LAB at the University of Bologna. The research is based on data acquired with the NASA/ESA HST at the Space Telescope Science Institute, which is operated by AURA, Inc., under NASA contract NAS5-26555.

**Author contributions:** F.R.F. designed the study and coordinated the activity. E.D., M.C. and S.R. analysed the photometric datasets. F.R.F, B.L. and E.D. wrote the first draft of the paper. G.B., A.M. and C.P. critically contributed to the paper presentation. All the authors contributed to the discussion of the results and commented on the manuscript.

**Competing Interests:** The authors declare no competing interests.

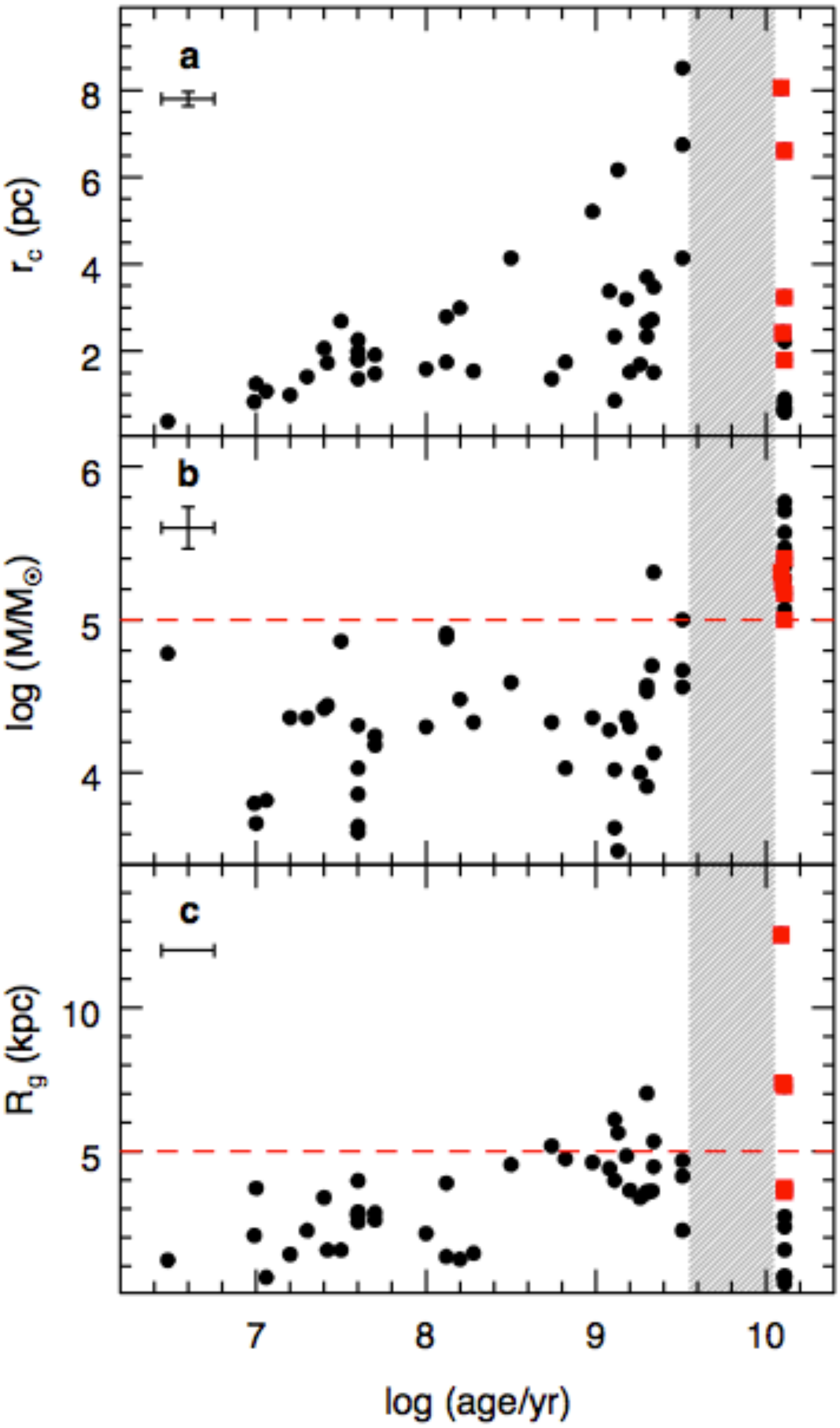

**Figure 1 - The drastic change in LMC cluster properties as a function of cosmic time.** Observed distribution of core radius[22] (rc), total mass[22] (M) and galactocentrinc distance[22] (Rg) versus chronological age[22] for the LMC GCs (panels a, b, and c, respectively). The average 1σ errors are marked in each panel. The ~10 Gyr long period of cluster formation quiescence[8-10] is marked with a grey shaded region. The dashed red lines in panels b and c mark, respectively, the limits in mass and galactocentric distance characterizing the recent cluster formation process. The five clusters discussed in this paper are shown as red squares.

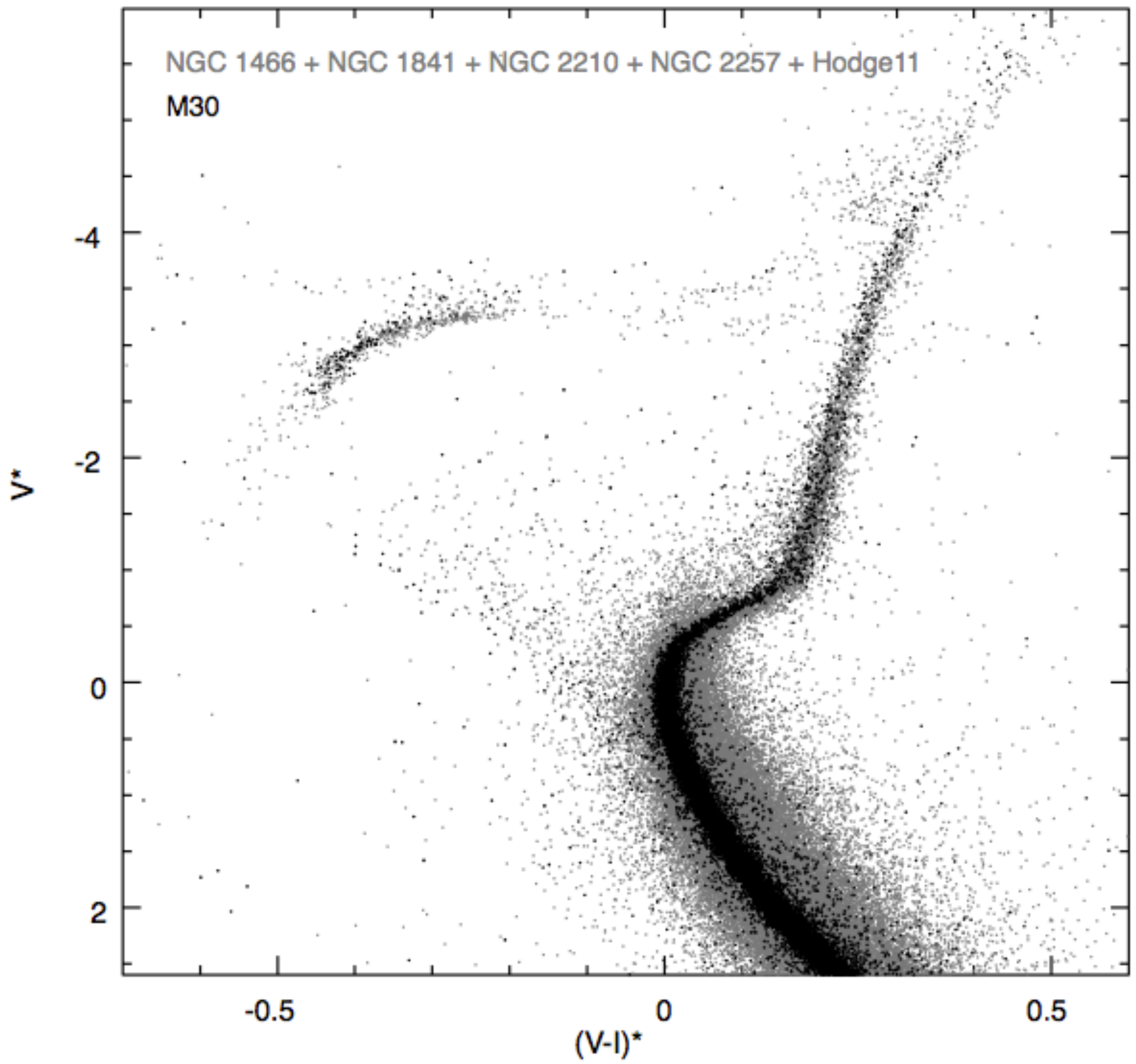

**Figure 2. The age of the 5 LMC clusters.** –The co-added n-CMDs of the 5 LMC clusters (in grey) is compared with that of the Galactic globular cluster M30[38] (t=13 Gyr)[39,40]. The comparison clearly demonstrates that the 5 clusters are all old and coeval, with an age of ~13 Gyr.

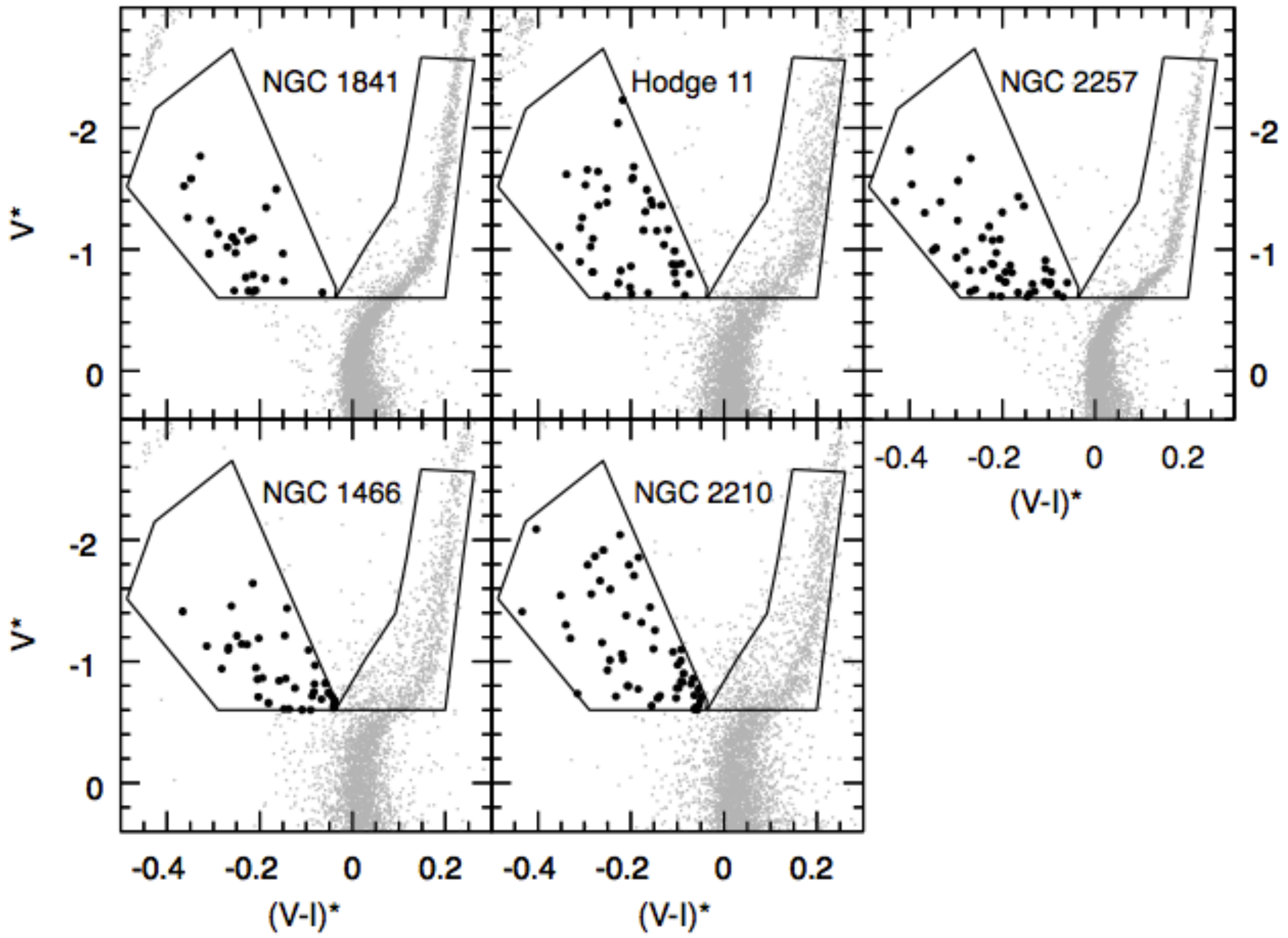

**Figure 3. Blue Straggler Star selection**. The selection boxes of BSSs and of the reference population are shown in the n-CMD of each cluster, where only stars measured within one half-mass radius are plotted. Only BSSs (black circles) and reference stars brighter than V* = −0.6 have been considered in the present analysis. For the two most contaminated clusters (Hodge 11 and NGC 2210) the selection box appears to be more populated on the red side, thus suggesting that this is the region where the field contamination is more severe (see Methods, Supplementary Figure 1 and Supplementary Table 1 for more details).

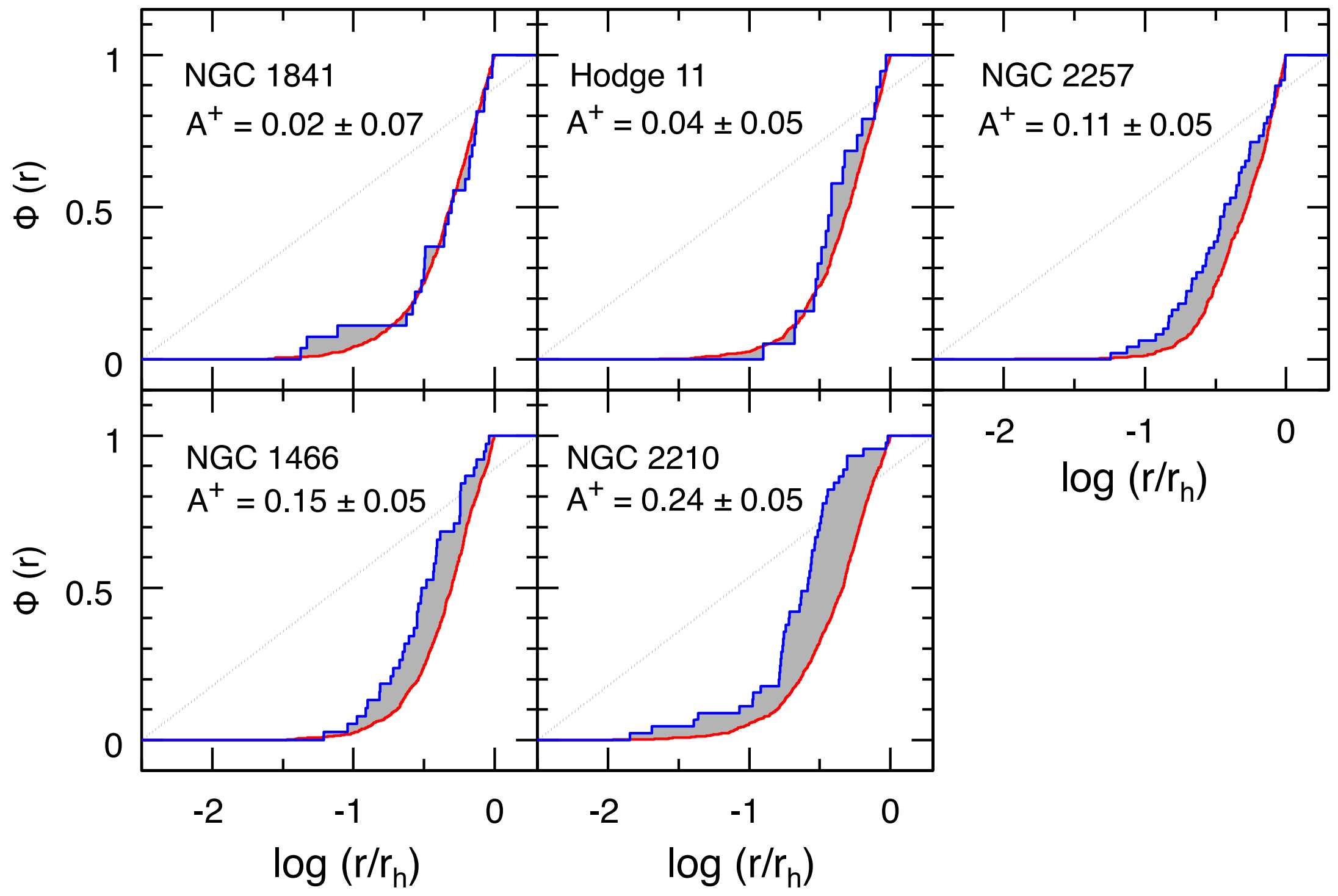

**Figure 4. Measure of the A+ parameter.** Cumulative radial distributions of BSSs (blue line) and reference stars (red line) in the five LMC GCs discussed in this paper. Only stars within one half-mass radius have been considered and the cumulative radial distributions are thus normalized to unity at $r_h$. The size of the area between the two curves (shaded in grey) corresponds to the labelled value of A+. Clusters are ranked in terms of increasing value of A+.

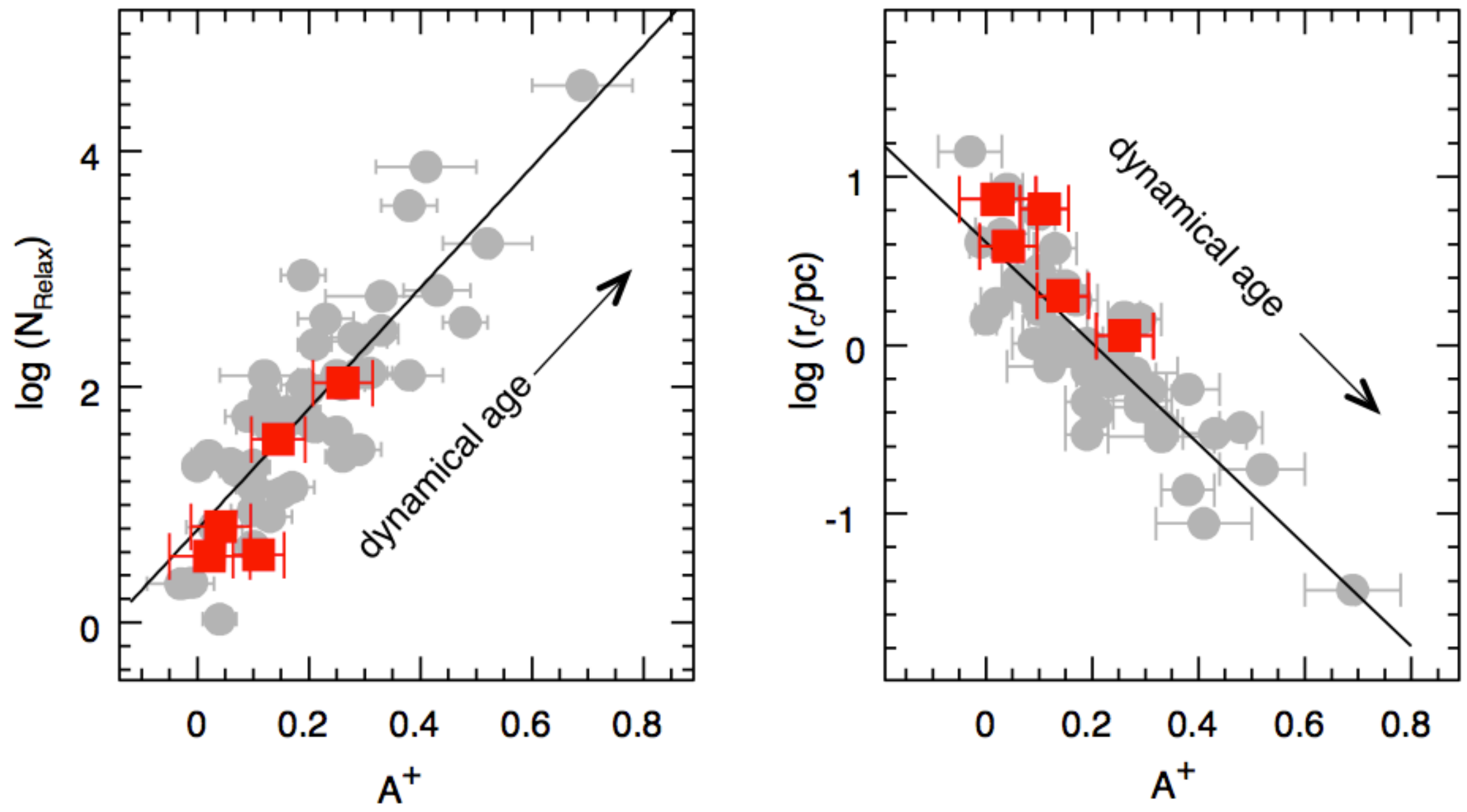

**Figure 5. Quantifying the internal dynamical evolution and its effect on the cluster physical size.** *Left Panel:* Relation between the segregation level of BSSs (measured by A+) and the number of current central relaxation times occurred since cluster formation ($N_{Relax}$) for the 5 LMC clusters studied here (large red squares) and 48 Galactic GCs (grey circles)[15]. The plotted 1σ errors have been computed as discussed in Methods. *Right Panel:* Relation between A+ and the core radius, illustrating that cluster sizes move toward smaller values with the long-term internal dynamical evolution of the system: compact clusters are dynamically more evolved than large-rc GCs.

**Table 1. Cluster parameters determined in this work. The quoted uncertainties are s.d.**

| Cluster | Gravitational centre | r.m.s. | offset from MG03 | $r_c$ [arcsec] [pc] | $r_h$ [arcsec] [pc] | $\log(t_{rc}/yr)$ | A+ |
|---|---|---|---|---|---|---|---|
| NGC 1466 | $03^h\ 44^m\ 32.72^s$ <br> $-71^o$ 40' 15.63" | 0.3" | 2.8" | $8.1^{+0.8}_{-0.7}$ <br> 2.0 pc | $24.5^{+0.2}_{-0.2}$ <br> 5.9 pc | 8.56 | 0.15 <br> ±0.05 |
| NGC 1841 | $04^h\ 45^m\ 22.49^s$ <br> $-83^o$ 59' 55.06" | 0.4" | 2.4" | $30.3^{+1.3}_{-1.3}$ <br> 7.3 pc | $57.9^{+3.9}_{-3.9}$ <br> 14.0 pc | 9.54 | 0.02 <br> ±0.07 |
| NGC 2210 | $06^h\ 11^m\ 31.69^s$ <br> $-69^o$ 07' 18.37" | 0.1" | 1.7" | $4.4^{+0.3}_{-0.5}$ <br> 1.1 pc | $15.9^{+0.6}_{-0.2}$ <br> 3.9 pc | 8.07 | 0.24 <br> ±0.05 |
| NGC 2257 | $06^h\ 30^m\ 12.59^s$ <br> $-64^o$ 19' 37.21" | 0.4" | 5.7" | $26.7^{+1.6}_{-1.4}$ <br> 6.5 pc | $56.4^{+2.6}_{-1.6}$ <br> 13.6 pc | 9.51 | 0.11 <br> ±0.05 |
| Hodge 11 | $06^h\ 14^m\ 22.99^s$ <br> $-69^o$ 50' 49.92" | 0.2" | 3.6" | $15.0^{+1.0}_{-0.8}$ <br> 3.6 pc | $36.3^{+1.5}_{-0.6}$ <br> 8.8 pc | 9.11 | 0.04 <br> ±0.05 |

# Supplementary Information

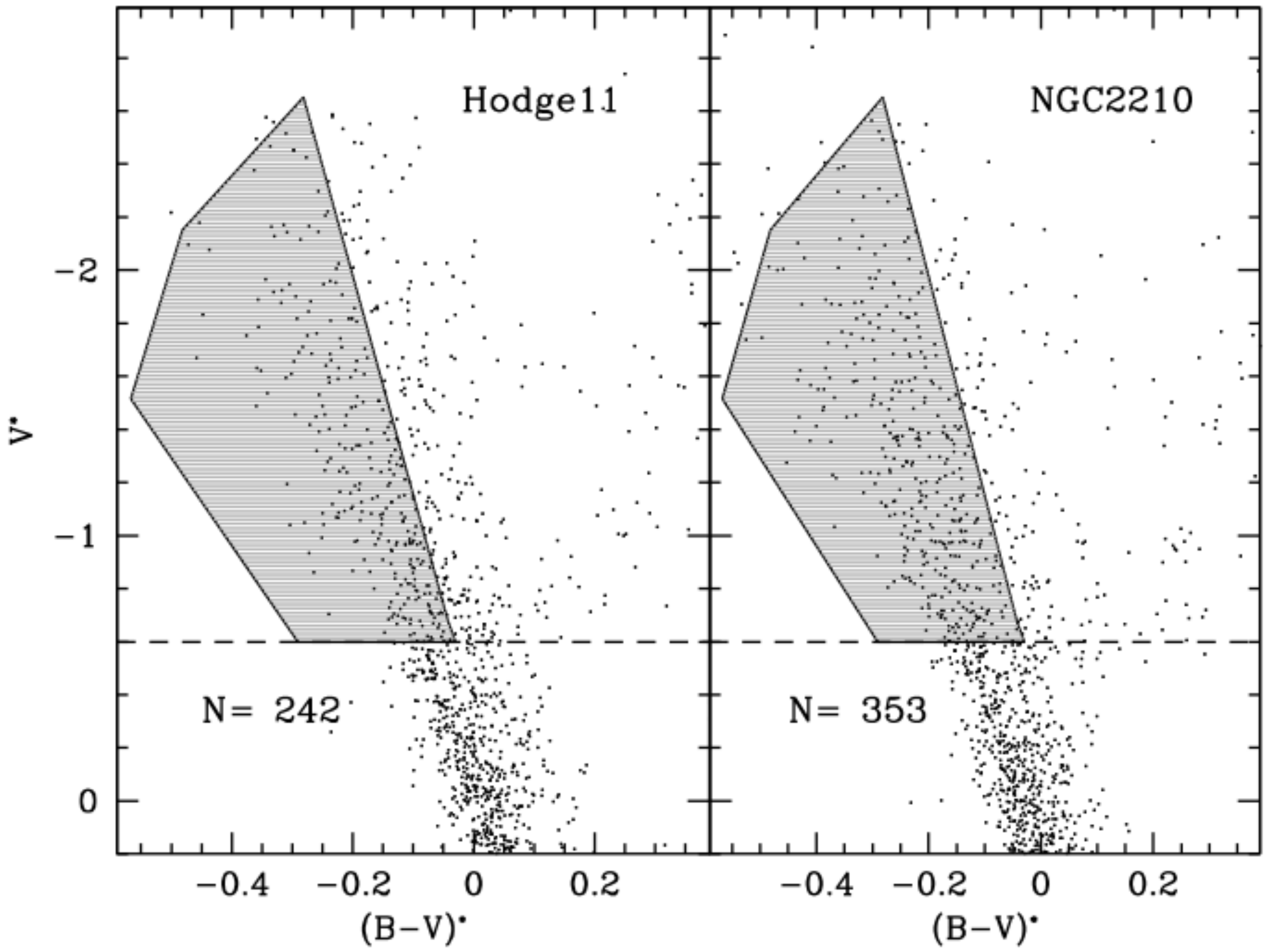

**Supplementary Figure 1. Field decontamination in the case of NGC 2210 and Hodge 11.** –The n-CMDs of the LMC fields adjacent to Hodge 11 and NGC 2210, zoomed in the BSS region, are shown. The grey shaded area marks the BSS selection box converted in the [V*, (B-V)*] n-CMD by using theoretical isochrones[48] of appropriate metallicity. The total number of contaminating stars (which preferentially populate the redder portion of the BSS selection box) are marked. When the considered cluster area (r<rh) is taken into account, they translate into 24 potential field interlopers for Hodge 11 and 6 for NGC 2210.

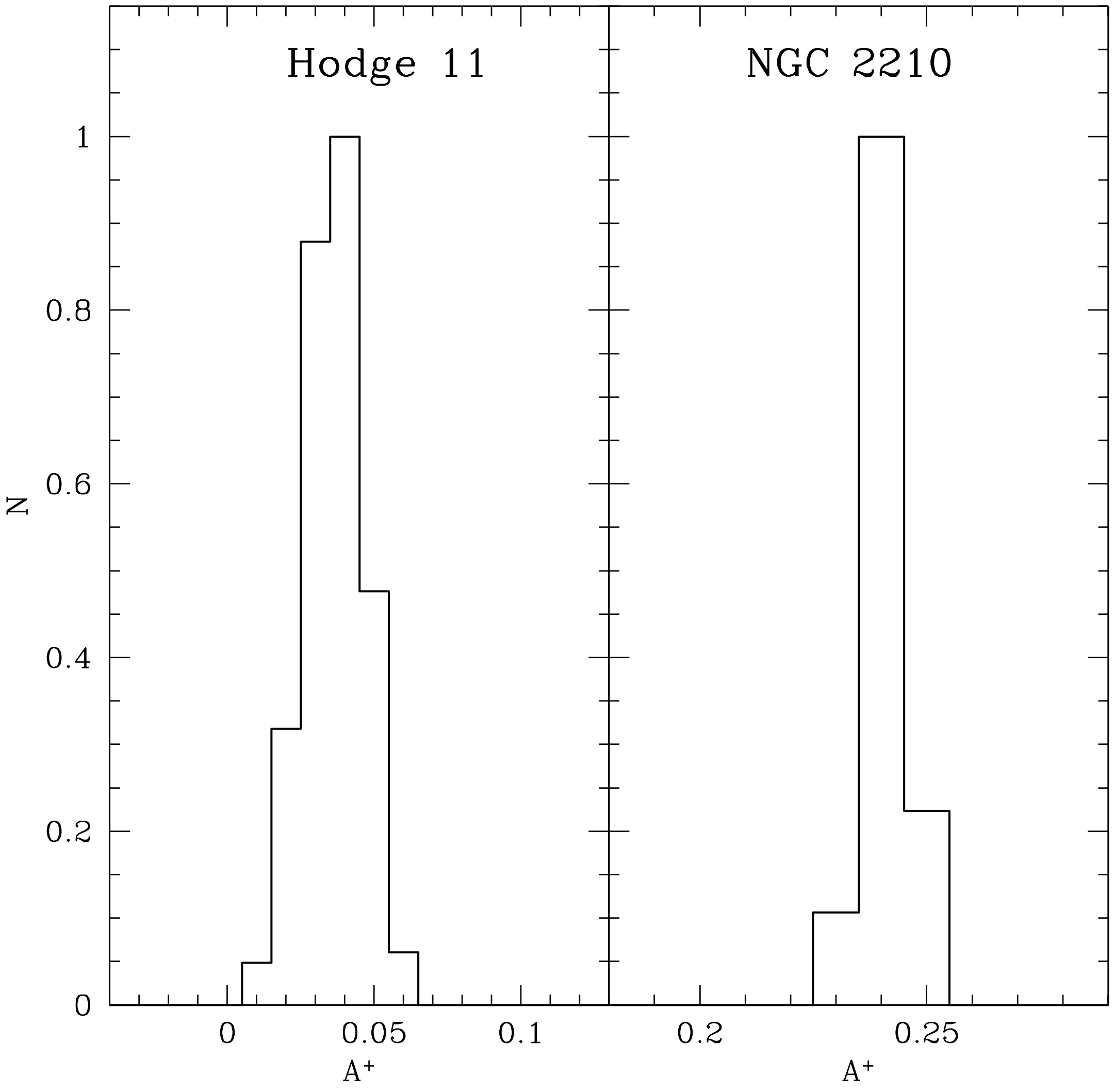

**Supplementary Figure 2. Field decontamination effect on the measure of A+.** – Normalized distribution of the values of A+ obtained from 5000 independent decontamination procedures applied to the BSS region of Hodge 11 and NGC 2210. In each realization, 24 (in Hodge 11) and 6 stars (in NGC 2210) have been randomly removed from the BSS region of the two clusters and the A+ parameter has been re-determined. As can be seen, in both cases the result is a peaked distribution with a well defined maximum (A+=0.04 for Hodge 11, and A+=0.24 for NGC 2210) and a small dispersion (σ=0.01).

**Supplementary Table 1. Field contamination.**

| Cluster | $N_{BSS}$ r<rh | $N_{field}$ 0< r<rc | $N_{field}$ rc< r<rh/2 | $N_{field}$ rh/2< r<rh |
|---|---|---|---|---|
| NGC 1466 | 38 | 0 | 0 | 0 |
| NGC 1841 | 27 | 0 | 0 | 0 |
| NGC 2210 | 52 | 0 | 1 | 5 |
| NGC 2257 | 49 | 0 | 0 | 0 |
| Hodge 11 | 45 | 4 | 1 | 19 |